\documentclass[twocolumn,showpacs,preprintnumbers,
               superscriptaddress,floatfix]{revtex4}
\usepackage{graphicx}
\usepackage{amsmath}
\usepackage{amssymb}

\renewcommand{\vec}[1]{\mbox{\boldmath $#1$}}

\newcommand{\e}{\rm{e}}                                 
\newcommand{\bra}[1]{\langle #1 \rvert}                 
\newcommand{\ket}[1]{\lvert #1 \rangle}                 
\newcommand{\bk}[2]{\langle #1 | #2 \rangle}            
 
\newcommand{\nn}{\nonumber}            
\newcommand{\Gm}{\mu}
\newcommand{\Gn}{\nu}

\newcommand{\Gg}{\gamma}
\newcommand{\Gd}{\delta}
\newcommand{\Ge}{\epsilon}

\newcommand{\Gl}{\lambda}
\newcommand{\Gt}{\theta}

\newcommand{\Btopi}{B\rightarrow\pi l\nu}
\newcommand{\Btorho}{B\rightarrow\rho l\nu}
\begin{document}

\preprint{YITP-04-32}
\preprint{HUPD-0403}

\title{A model independent determination of $|V_{ub}|$ 
using the global $q^2$ dependence of the dispersive bounds on the 
$B\rightarrow\pi l\nu$ form factors}

\author{Masaru Fukunaga}
\affiliation{Department of Physics,
             Hiroshima University, Higashi-Hiroshima, 
             Hiroshima 739-8426, Japan}
\author{Tetsuya Onogi}
\affiliation{Yukawa Institute for Theoretical Physics,
             Kyoto University, Kyoto 606-8502, Japan}


\pacs{12.38.Gc (temporary)}

\begin{abstract}
We propose a method to determine the CKM matrix element $|V_{ub}|$ 
using the global $q^2$ dependence of the dispersive bound 
on the form factors for  $B\rightarrow \pi l\nu$ decay.
Since the lattice calculation of the $B\rightarrow \pi l\nu$ form factor 
is limited to the large $q^2$ regime, only the experimental 
data in a limited kinematic range can be used in a conventional 
method. In our new method which exploits the statistical 
distributions of the dispersive bound proposed by Lellouch, we can 
utilize the information of the global $q^2$ dependence for all
kinematic range.
As a feasibility study we determine $|V_{ub}|$ by combining the form factors 
from quenched lattice QCD, the dispersive bounds, and the experimental
data by CLEO. We show that the accuracy of $|V_{ub}|$ can be improved
by our method. 
\end{abstract}

\maketitle

\section{Introduction}
\label{sec:Introduction}
Precise determination of the Cabibbo-Kobayashi-Maskawa (CKM) matrix 
elements from the B, D, and K decays is one of the major goals in 
flavor physics. 
By measuring both the sides and the angles of the unitarity triangle, 
one can test the consistency of the standard model, which can either 
verify the standard model or probe a signal of new physics. 
Although useful for the consistency check against  $\sin(2\phi_1)$, 
the CKM matrix element $|V_{ub}|$ is one of the most poorly 
known quantities at present. It would be important to reduce the 
uncertainties further.

There are two ways to determine $|V_{ub}|$, i.e. the determination 
from the inclusive semileptonic decay $B \rightarrow X_u l \nu$ , 
and the determination from the exclusive semileptonic decay 
$\Btopi$ or $\Btorho$.
Since both methods suffers from different systematic errors 
from the experiment and the theory, having independent 
results from the inclusive and the exclusive processes
are necessary for the reliable determination of $|V_{ub}|$.

In the former method,  the hadronic matrix element
in the inclusive process is the operator product expansion (OPE) 
in $1/m_b$. It is suggested by Bauer {\it et al.}~\cite{Bauer:2001rc}
that with an 
appropriate choice of the combined kinematical cut ($m_X^2$  and $q^2$) 
to remove the charm background, theoretical errors from 
various higher order corrections in OPE are controlled so that 
$|V_{ub}|$ can be determined at the level of around 10\% or below. 

The latter method requires the computations of the form factors 
which describe the hadronic weak matrix elements for the exclusive 
processes. Lattice QCD is a promising tool for this purpose. 
However at large recoil, the cutoff effects of order $O(a E)$ 
becomes non negligible 
where $a$ is the lattice spacing and $E$ is the recoil energy of 
the daughter mesons such as $\pi$ or $\rho$. 
Therefore the precise computation of the form factor is limited only 
for large $q^2$ region in the lattice QCD method. 
Although B factory experiments can measure the differential decay rate 
at high $q^2$ regime, they are statistically limited while 
the rich experimental data for the low $q^2$ region 
will remain untouched. It would be ideal if one could 
make a precise prediction of the semileptonic decay form factors 
for the whole $q^2$ range. 

Lellouch proposed a statistical method in which one uses the form 
factor values with theoretical errors for $\Btopi $
at high $q^2$ from lattice QCD to give a distribution of the
dispersive bound for the form factors for the small $q^2$ 
region~\cite{Lellouch:1995yv}. 
Although the idea is attractive, with his original idea one could 
only obtain a loose bound for the form factor. 
Later several authors pointed out that this bound can be 
improved with additional inputs near $q^2_{max}$
~\cite{Becirevic:vw,Mannel:1998kp}.
Recently CLEO gave the $q^2$ dependence of the $\Btopi$ 
decay~\cite{Athar:2003yg}.
Although the overall normalization for the form factor is unknown 
up to $|V_{ub}|$, the CLEO results 
give a strong constraint on the $q^2$ dependence of the $\Btopi$ 
form factor. In this paper, we propose to use the information on the 
$q^2$ dependence of the $\Btopi$ from the experiment 
to restrict the statistical distribution of the dispersive bound.
We show that this additional input leads to a better determination of 
$|V_{ub}|$.

This paper is organized as follows.
In \ref{sec:Basic}, we explain our basic idea for 
$|V_{ub}|$ for determination using lattice data, dispersive bounds,
and the experimental data. The explicit method how to combine 
those data explained in detail in section~\ref{sec:Method}. 
In section~\ref{sec:Results},  our results on the improved bound 
for the form factor as well as $|V_{ub}|$ are presented.
In section~\ref{sec:Errors},  we discuss the systematic errors. 
We summarize our results in section \ref{sec:Conclusion}.
The appendix is devoted to a review of dispersive bound.

\section{Basic Idea}
\label{sec:Basic}
The matrix element of the heavy-to-light semileptonic decay
$B\rightarrow \pi l\nu$ is parameterized as
\begin{eqnarray}
  \bra{\pi^+(k)}V^\Gm(0)\ket{{B}^0(p)}
& = &\left(p+k-q\frac{m_B^2-m_\pi^2}{q^2}\right)^\Gm f^+(q^2)
\nn\\
& & +q^\Gm\frac{m_B^2-m_\pi^2}{q^2}f^0(q^2),
\label{eqff}
\end{eqnarray}                                 
where $V^\Gm=\bar{q}\Gg^\Gm b$ and $q^2$
ranges from $q^2_{min}=m_l^2 (\sim 0)$ GeV$^2$ to
$q^2_{max}=(m_\pi-m_l)^2$. 
The differential decay rate is written as
\begin{equation}
  \label{eqBtopi_differential_decay_rate}
  \frac{d\Gamma(B\to\pi l\nu)}{dq^2} = 
  \frac{G_F^2}{24\pi^3}
  |V_{ub}|^2  [(v\cdot k)^2-m_\pi^2]^{3/2} 
  |f^+(q^2)|^2.
\end{equation}

Since the discretization error becomes uncontrollable for the
momentum much larger than $\Lambda_{\mathrm{QCD}}$, the
calculation of the form factors is feasible only in the
large $q^2$ (small recoil) region ($q^2\gtrsim$ 16~GeV$^2$),
where the spatial momentum of pion is lower than roughly
1~GeV. The CKM matrix element $|V_{ub}|$ can be obtained by
combining the experimental data integrated above some $q^2$
value, and the lattice results for the form
factor $|f^+(q^2)|^2$ integrated in the same region with an
appropriate kinematical factor. However, due to the limited 
statistics the experimental measurement for this energy range 
still has a large error. Although in principle there is 
nothing wrong with this method, one disadvantage is that 
the experimental data for the rest of the kinematic range
with much better statistics cannot be used.  

An alternative way is to extrapolate the form factor 
to all kinematic range and combine with the full experimental
data. For this purpose,  we need to have the following three 
ingredients 
\begin{enumerate}
\item the experimental data of the partial decay rate 
      $\int_{q_i^2}^{q_{i+1}^2} dq^2 d\Gamma/dq^2$ 
      for the kinematic ranges $q_i^2 \le q^2 \le q_{i+1}^2$, 
\item the lattice results on the form factor $f^+(q^2)$ for
large $q^2$ region,
\item a reliable method to extrapolate the form factor to 
lower $q^2$ region. 
 \end{enumerate}
In order to avoid model dependences, we exploit the dispersive 
bound to extrapolate the form factor. The dispersive bound 
is an exact bound on the form factors $f^0(q^2)$ and $f^+(q^2)$ 
for all kinematic range of 
$q^2$~\cite{Okubo:ih,Bourrely:1980gp,deRafael:1993ib,Boyd:1994tt}. 
This is derived from the 
dispersion relation and the operator product expansion (OPE) 
of the two-point correlation function of the heavy-light (bu) 
current in deep Euclidean region. 
If we have an additional information on the form factors $f^0, f^+$ 
at some physical 
kinematic points $q_{i_1}^2, \cdots , q_{i_N}^2$ (from  
the lattice calculation, for instance),  the bound can be improved further. 
In this case, the upper and the lower bounds are the solutions 
of the quadratic equations whose coefficients are determined 
by the lattice inputs as well as other inputs such as OPE results.
For a set of lattice data $\vec{f}=\{\vec{f^+}, \vec{f^0} \}$
where $\vec{f^+} \equiv \{ f^+(q_1^2), \cdots f^+(q_N^2)\}$ 
and $\vec{f^0} \equiv \{ f^0(q_1^2), \cdots f^0(q_N^2)\}$, 
the upper and lower bounds are uniquely determined 
as a function of $q^2$
as long as the solution of the quadratic equation exists.
Let us denote the bounds as
\begin{equation}
F_{up/lo}^+(q^2;\vec{f}), F_{up/lo}^0(q^2;\vec{f}).
\end{equation}
 \begin{figure}[here]
  \begin{center}
      \includegraphics[width=9cm]{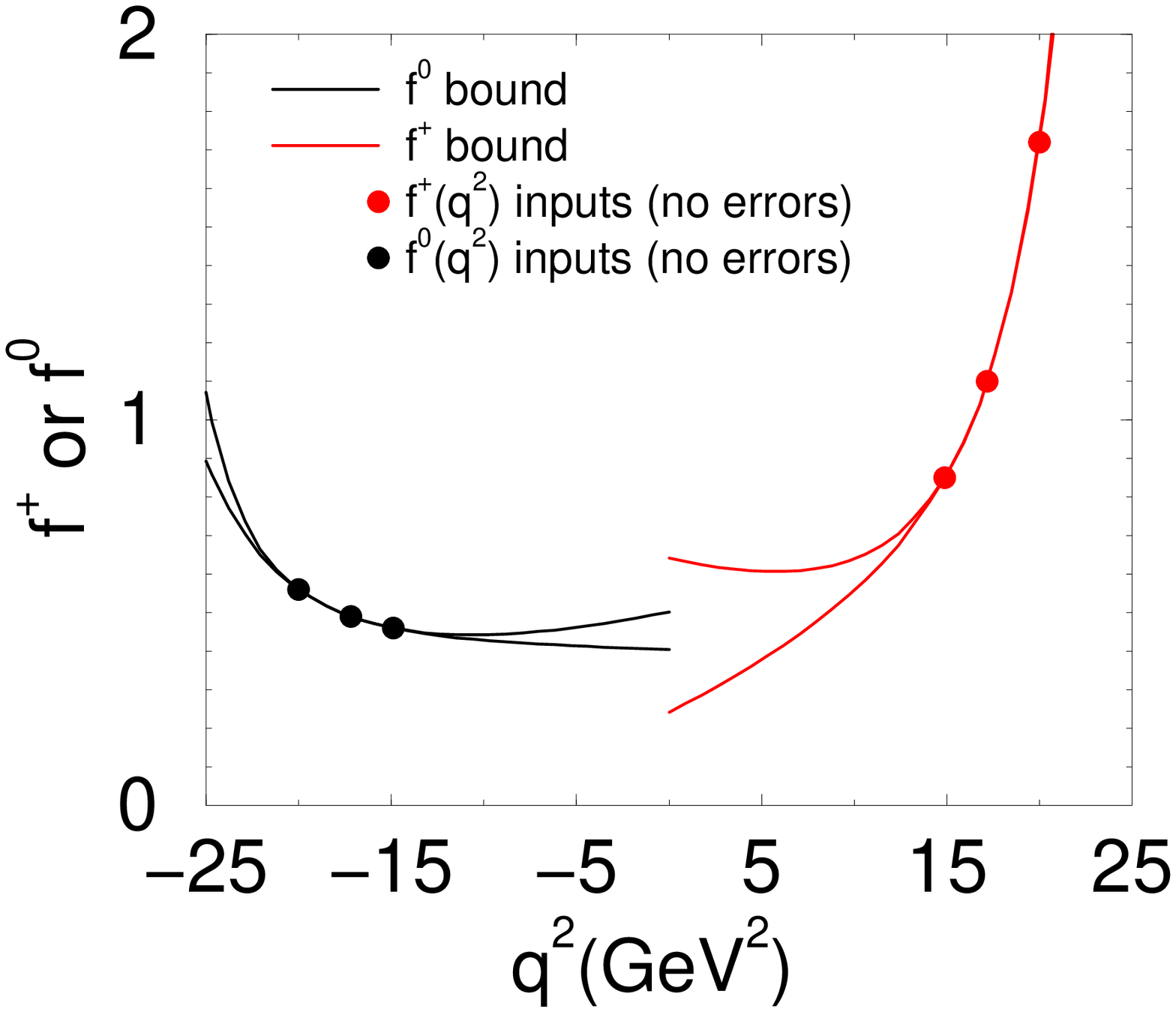}\\
      \includegraphics[width=9cm]{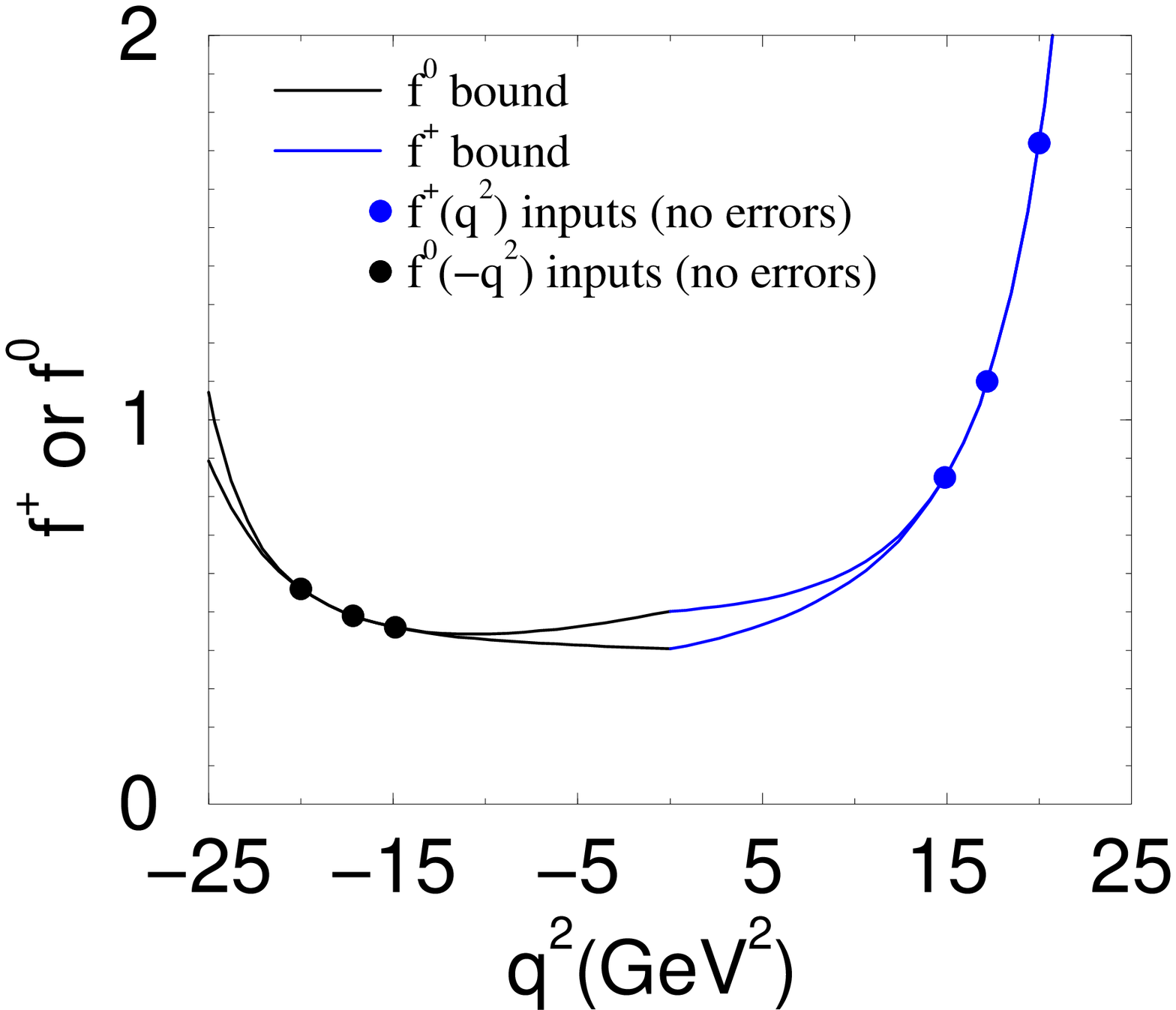} 
  \end{center}
   \caption{bound of $f^+(t),f^0(-t)$ for one example set of inputs $\vec{f}$}
    {\footnotesize The top figure shows the independent bounds for 
     $F^0_{up/lo}$ and $F^+_{up/lo}$ 
     and the bottom figure shows the bounds with the kinematical
     constraint $f^+(0)=f^0(0)$ 
   } 
     \label{fig:disp_no_error}
 \end{figure}
Fig.~\ref{fig:disp_no_error} shows 
a typical shape of the dispersive bound using a mock data 
and it is suggested that the inputs at three points 
could give rather stringent constraints.

One problem is that in practice the additional inputs from the 
lattice calculation always has some theoretical errors so that 
this bound has uncertainties. Lellouch proposed a statistical 
treatment and derived a probability distribution the dispersive bound
~\cite{Lellouch:1995yv}. 
He made a random sample set of form factors which obey
the Gaussian distributions where the central value and the 
deviations are taken from the central values and the error of 
the lattice data. For each sample set,  
the quadratic equations which determine the upper and lower bounds 
are solved.  He imposed the following condition,
\begin{description}
\item[ Condition A] (consistency condition):
\begin{itemize}
\item The quadratic equations to determine the upper/lower bounds 
      should have real solutions.
\item The solutions of the upper/lower bounds 
      should allow the kinematical condition
      $f^+(0)=f^0(0)$.
\end{itemize}
\end{description}
The sample is accepted when condition A is satisfied 
and discarded otherwise. 
Then he makes a distributions of the upper and lower
bounds from only the accepted solution, which are 
conditional distributions. 

\begin{figure}[here]
\resizebox{80mm}{!}{\includegraphics[angle=-90]{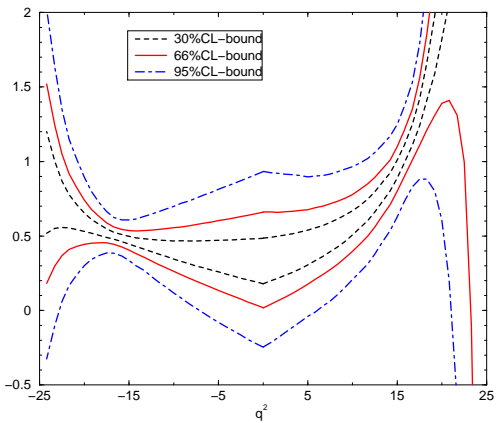}}
\caption{The confidence level bound for $f^+(t)$, $f^0(-t)$. 
The input data is lattice data(Lellouch).}
\label{fig:disp_CL}
\end{figure}

As shown in Fig.~\ref{fig:disp_CL}, the confidence level bound 
at $q^2\sim 0$ and $q^2=q^2_{max}$ has a huge spread.  
Although the dispersive bound method gives bounds at certain confidence
levels, they are not strong enough to constrain the CKM matrix element 
$|V_{ub}|$ with high accuracy. 

In the following we would like to consider how to make 
the best use of the experimental data and the theoretical 
results. The large width of the dispersive bound arises 
because the statistical bound was obtained at each $q^2$ 
without considering the correlation in $q^2$. 
However, as shown in Fig. 1, each sample of the 
dispersive bound from the Monte Carlo sample 
based on the lattice data gives a very stringent bound 
and the $q^2$ dependence can be obtained with, say, 10\% 
accuracy. On the other hand, from CLEO experiment 
we also know the $q^2$ dependence up to an overall normalization 
from unknown factor $|V_{ub}|$.  
The knowledge of the  $q^2$ dependence of the form factors from 
both the theory and experiment can be an additional input 
for the determination of $|V_{ub}|$. 
To extract $|V_{ub}|$, instead of combining the experimental 
data and the lattice results for a single $q^2$ bin, 
we make a kind of simultaneous fit for all the $q^2$ bins. 
Precisely speaking, since the lattice results are transformed 
into the distributions of the upper and lower bounds, a little 
modifications are necessary, namely rather than fitting $|V_{ub}|$
we obtain a conditional distributions of $|V_{ub}|$ 
using all information from the experimental and the lattice results.
In other words, we can reweight the statistical distribution 
of the form factor bounds 
using the $q^2$ distribution from the 
experiment as an additional input, which we will explain 
more in detail in the next section. 

We also note that the soft pion theorem 
can also be used for additional inputs.
The soft pion predicts that the form factor 
near the zero pion recoil limit ($q^2 \rightarrow q^2_{max}$),
the form factor behaves as 
 \begin{eqnarray}
f^0(q^2_{max})& \rightarrow &\frac{f_B}{f_\pi}\label{eq3-1}\\
f^+(q^2_{max})& \rightarrow &\frac{g f_{B^*}}{f_\pi}\frac{1}{1-q^2/m_{B^*}^2},
\label{eq3-2}
\end{eqnarray}
where $g$ is the $B^* B\pi$ coupling and $f_B \sim f_{B^*}$ 
are the decay constants of $B$ and $B^*$ mesons~\cite{Burdman:1993es}
. 
Therefore provided that $g$, $f_B$, $f_{B^*}$ are known, 
the soft pion relation can be used to gives additional inputs 
of the form factor at $q^2_{max}$ for the dispersive bounds.

\section{Method}
\label{sec:Method}
In the previous section, we explained the basic idea for 
extracting the CKM element $|V_{ub}|$ by 
combing the lattice results, experimental data, and the dispersive
bound. In this section we explain our method more in detail.

Due to limited statistics and limited energy resolution in $q^2$ etc, 
usually the whole kinematic range is divided into $N_{bin}$ bins 
($q^2_i < q^2 < q^2_{i+1}, i = 1, \cdots, N_{bin}$ ) and 
we only know the partial decay rates for those $q^2$ bins 
\begin{equation}
\Gamma^{exp}_i\equiv\int_{q^2_i}^{q^2_{i+1}} dq^2 \frac{d\Gamma}{dq^2}
 = 
\frac{G^2|V_{ub}|^2}{192\pi^3m_B^3}
\int_{q^2_0}^{q^2_1} dq^2 |f^+(q^2)|^2  \lambda(q^2)^{3/2}.
\label{eqGamma_i}
\end{equation}
This partial decay rates are indeed measured
by CLEO collaboration~\cite{Athar:2003yg}.

By integrating the dispersive bounds of the form factor 
$F^+_{up/lo}(q^2;\vec{f})$ , 
we can predict the upper and lower bounds of the partial 
decay rate over certain $q^2$ bins 
up to the overall factor $|V_{ub}|^2$.
\begin{eqnarray}
&& \gamma^{up/lo}_i(\vec{f})
\nn\\
&\equiv &
\frac{G^2}{192\pi^3m_B^3}\int^{t_{i+1}}_{t_i}dt 
\left|F^+_{up/lo}(t;\vec{f}^+,\vec{f}^0) 
\right|^2\Gl(t)^{3/2},
\label{eqgamma}
\end{eqnarray}
where the theoretical upper/lower bounds of the partial decay rates 
are $\Gamma^{up/lo}_i(\vec{f})=
|V_{ub}|^2 \times \gamma^{up/lo}_i(\vec{f})$ . 

Let us now explain our methods, which is composed of five steps.
\begin{enumerate}
\item We generate samples of $\vec{f}$ from the Gaussian 
distribution with central values and errors from the lattice 
data. The samples of $B^* B\pi$ coupling $g$ and $f_B\sim f_{B^*}$ 
are also generated from the some distributions based on the 
best knowledge of $g$ and $f_B$, from which 
the samples of $\vec{f}_{max}\equiv\{f^+(q^2_{max}), f^0(q^2_{max})\}$ 
are produced.
They can be used as additional data of the form factor 
for the dispersive bounds. 
\item From each sample of $\{ \vec{f}, \vec{f}_{max} \}$, 
the dispersive bounds $F^+_{up/lo}(q^2;\vec{f},\vec{f}_{max})$ 
(and $F^0_{up/lo}(q^2;\vec{f},\vec{f}_{max})$) are derived and 
each sample is accepted or rejected under the Condition A,  
so that a conditional distribution is produced.
\item For each accepted sample, we compute the upper and lower bounds 
of the partial decay rate $\gamma^{up/lo}_i(\vec{f},\vec{f}_{max})$'s 
for a given set of $q^2$ bins.
\item We create  samples  of $|V_{ub}|$ and $\Gamma^{exp}_i$'s.
We assume $|V_{ub}|$  distributes uniformly within a conservatively 
wide range, i.e.  $|V_{ub}|=[1\times 10^{-3}, 6\times 10^{-3}]$. 
and $\Gamma^{exp}_i$'s  are distributed by a Gaussian around their central 
values with variances given by the CLEO data. 
\item We further impose the following physical condition for the 
set of samples $\{\vec{f},g,f_B,|V_{ub}|,\Gamma^{exp}_i\}_{CondA}$.
\begin{description}
\item[ Condition B] (physical condition):
\begin{itemize}
\item  The experimental data $\Gamma^{exp}_i$'s should 
 lie within the upper and lower bounds from the theory 
 simultaneously for all $i$ , i.e.\\
$|V_{ub}|^2 \gamma^{lo}_i < \Gamma^{exp}_i 
< |V_{ub}|^2 \gamma^{up}_i$ (i=$1,\cdots,N_{bin}$).
\end{itemize}
\end{description}
Try this test for all samples, 
select only those combinations which satisfies the 
condition B and reject others. 
This makes another conditional distribution 
of the set of samples 
$\{\vec{f},g,f_B,|V_{ub}|,\Gamma^{exp}_i\}_{CondAB}$ 
which satisfies both Condition A and B.
\end{enumerate}


Mathematically the original probability for the 
set of samples $\{\vec{f},g,f_B,|V_{ub}|,\vec{\Gamma}^{exp}\}$
is given by the following product
\begin{eqnarray}
&& P(\vec{f},g,f_B,|V_{ub}|,\vec{\Gamma}^{exp})
\nn\\
&=&
N \mathcal{P}_{\vec{f}}(\vec{f}) \mathcal{P}_g(g) 
\mathcal{P}_{f_B}(f_B)
\mathcal{P}_{exp}(\vec{\Gamma}^{exp})
\mathcal{P}_{|V_{ub}|}(|V_{ub}|), 
\label{org_prob}
\end{eqnarray}
where N is the normalization constant. 
The probability  $\mathcal{P}_{\vec{f}}, \mathcal{P}_{f_B}, 
\mathcal{P}_{exp}$ are Gaussian distributions based on theoretical 
or experimental data, where as $\mathcal{P}_g, \mathcal{P}_{|V_{ub}|}$ 
are uniform distributions. 

After imposing the Condition A, 
the probability becomes 
\begin{eqnarray}
&& P_A(\vec{f},g,f_B,|V_{ub}|,\vec{\Gamma}^{exp})
\nn\\
&=&
\left\{ 
\begin{array}{@{\,}ll}
N' P(\vec{f},g,f_B,|V_{ub}|,\vec{\Gamma}^{exp}) 
       &  \mbox{(if Cond A is satisfied), }  \\
    0  &  \mbox{(otherwise),}
\end{array}
\right. 
\nonumber
\label{condi_prob_A}
\end{eqnarray}
where $N'$ is another normalization constant to make the total 
probability unity. 
By further imposing the Condition B, the probability then becomes 
\begin{eqnarray}
&& P_{AB}(\vec{f},g,f_B,|V_{ub}|,\vec{\Gamma}^{exp})
\nn\\
&=&
\left\{ 
\begin{array}{@{\,}ll}
N'' P(\vec{f},g,f_B,|V_{ub}|,\vec{\Gamma}^{exp})
           & \mbox{(if Cond A+B is satisfied),} \\
    0      & \mbox{(otherwise),} 
\end{array}
\right.
\nonumber
\label{condi_prob_AB}
\end{eqnarray}
where $N''$ is  another normalization constant.

Using this last conditional distribution $P_{AB}$, 
the distributions of any quantities defined 
from the set of samples 
$\{\vec{f},g,f_B,|V_{ub}|,\vec{\Gamma}^{exp}\}_{AB}$, 
can be derived, which of course includes  that of $|V_{ub}|$.

\section{Results}
\label{sec:Results}
\subsection{The setup}
In this section we explain our results based on the lattice 
results and experimental data. The experimental data 
are taken  from the results from  CLEO collaboration~\cite{Athar:2003yg}.
Table~\ref{tab:CLEO} shows the partial branching fraction 
$\Gamma_i/\Gamma_{total}$. 
\begin{table}[here]
 \begin{center}
  \begin{tabular}{|c|c|}\hline        
  \multicolumn{1}{|c|}{$q^2$bin}    
  & $[(\int d\Gamma/d q^2)*dq^2]/\Gamma_{total}$\\ \hline
  \multicolumn{1}{|c|}{$0\le q^2 \le 8GeV^2$}     
  & $(0.43\pm 0.11)\times 10^{-4}$  \\
  \multicolumn{1}{|c|}{$8GeV^2\le q^2 \le 16GeV^2$} 
  & $(0.65\pm 0.11)\times 10^{-4}$  \\
  \multicolumn{1}{|c|}{$16GeV\le q^2 \le q^2_{max}$} 
 & $(0.25\pm 0.09)\times 10^{-4}$ \\ \hline
  \end{tabular}
  \caption{CLEO data of partial branching ratio for $B\rightarrow \pi
  l \nu$}\label{tab:CLEO}
 \end{center}
\end{table}
In order to convert $\Gamma_i$ we take $\Gamma_{total}=[4.29\pm 0.04 ] 
\times10^{-13} GeV$ ~\cite{ref:PDG}. 
There are lattice calculations from four different lattice
groups in the quenched 
approximation~\cite{Bowler:1999xn,Abada:2000ty,
El-Khadra:2001rv,Aoki:2001rd}.
In this section, we take the lattice result from JLQCD 
collaboration~\cite{Aoki:2001rd}.
We picked the form factor data at three points of $q^2$'s for our analysis, 
which are listed in Table~\ref{tab:JLQCD}.
\begin{table}[here]
 \begin{center}
  \begin{tabular}{|c|c|c|}           \hline
   \multicolumn{1}{|c|}{$q^2(GeV^2)$} & $f^+(q^2)$     
   & $f^0(q^2)$    \\ \hline
   \multicolumn{1}{|c|}{$17.79$}       & $1.03\pm 0.22$ 
   & $0.407\pm0.092$ \\
   \multicolumn{1}{|c|}{$19.30$}       & $1.24\pm 0.21$ 
   & $0.45\pm0.11$ \\
   \multicolumn{1}{|c|}{$20.82$}       & $1.54\pm 0.27$ 
   & $0.51\pm0.14$ \\ \hline
  \end{tabular}
  \caption{lattice results of $f^+(q^2)$ and $f^0(q^2)$ by JLQCD}
\label{tab:JLQCD}
 \end{center}
\end{table}
The recent $B^*B\pi$ coupling $g$ in quenched lattice QCD gives 
$g =0.48\pm 0.03 \pm 0.11$ for the static limit and 
$g =0.58\pm 0.06 \pm 0.10$ for the b quark 
from the quenched lattice calculations~\cite{Abada:2003un}. 
The QCD sum rule gives $g = 0.36 \pm 0.10$
~\cite{Khodjamirian:1999hb}. The predictions from 
quark models are in the range $0.3 < g < 0.8 $ \cite{quark_models}.
In our analysis
we took uniform distribution $g = [ 0.3,0.9 ]$.
The decay constant $f_B$ is known with 30\% error.
Since this error is much larger than the splitting of 
$f_{B^*}$ and $f_B$, we took $f_{B^*}=f_B= 190 \pm 30$ MeV 
for simplicity, taking the best estimate by the lattice 
calculation~\cite{Kronfeld:2003sd, Aoki:2003xb}.

In step 1 and 4, based on the above information  we generated 
${\cal O}(10^7)$ set of  samples  for $\{\vec{f},g,f_B\}$
, 2,000 set of samples for $\{ \vec{\Gamma}^{exp}\}$, 
and 2,000 samples for $|V_{ub}|$.

\begin{figure}[here]
\begin{center}
\resizebox{80mm}{!}{\includegraphics[angle=-90]{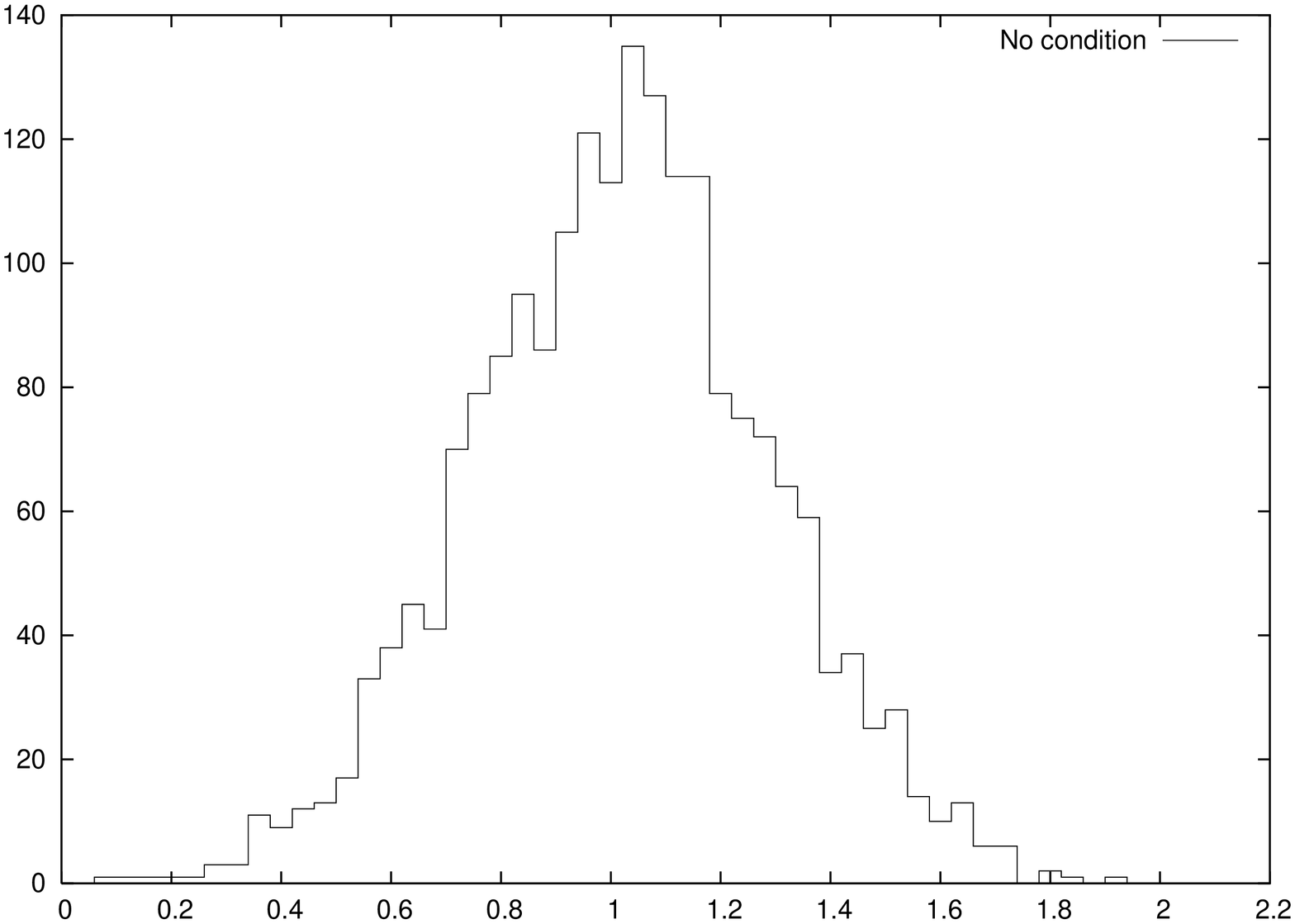}}
\resizebox{80mm}{!}{\includegraphics[angle=-90]{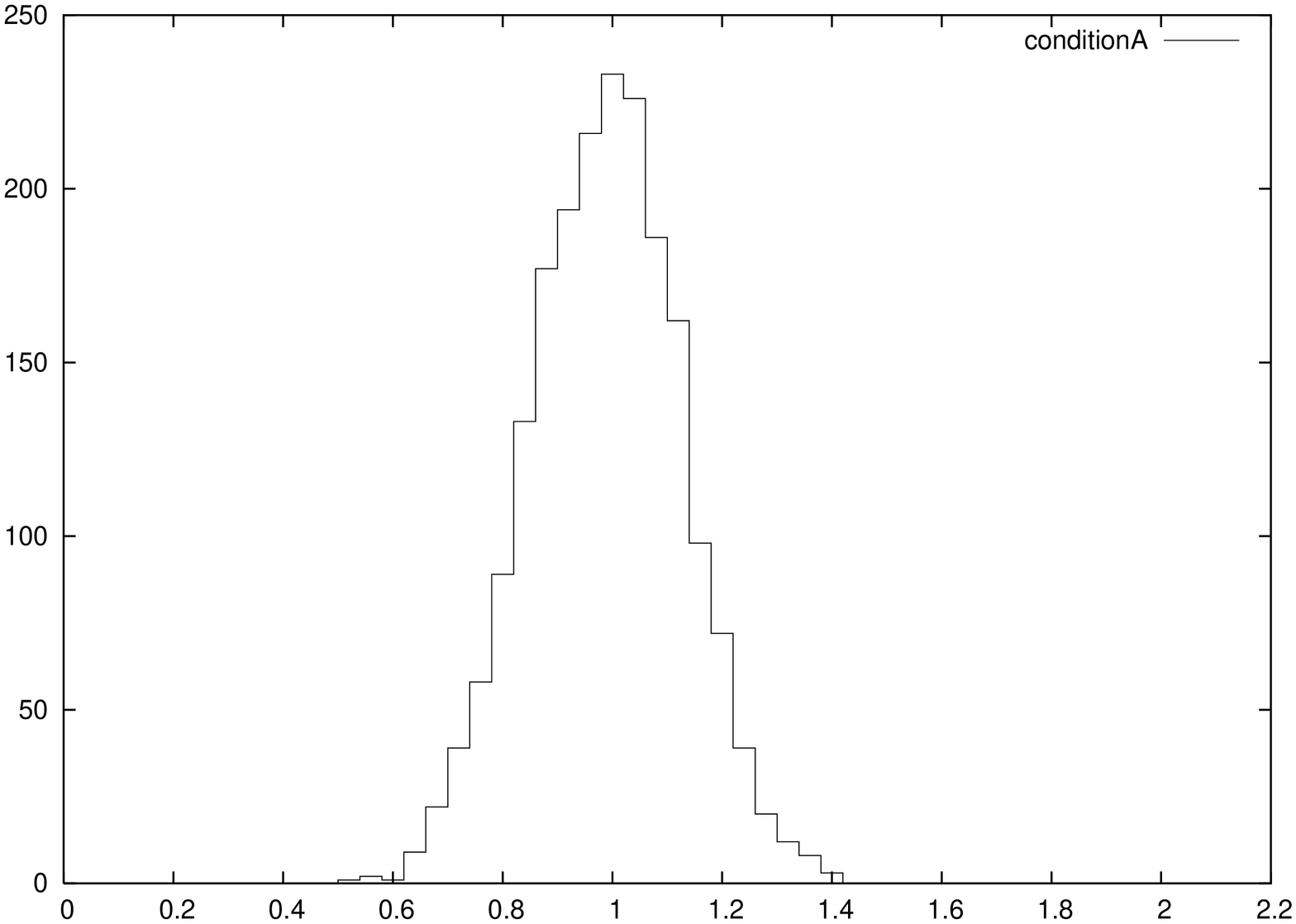}}
\caption{distribution of form factor $f^+(q^2=17.79GeV^2)$. 
The Top figure is the original samples from Gaussian distribution and 
the bottom is the samples which survived condition A.}
\label{fig:histvecf}
\end{center}
\end{figure}

Imposing the condition A in step 2, 2000 set of samples 
of $\{\vec{f},g,f_B\}$ survives.   In Fig.~\ref{fig:histvecf}, 
the histogram for the samples of $\vec{f}$ is shown. 
We find that the dispersive bound actually reduces 
the variance of the lattice data significantly.
%
%

\begin{figure}[here]
\resizebox{80mm}{!}{\includegraphics{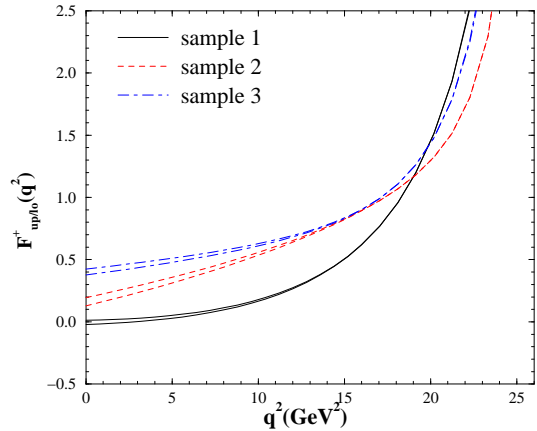}}
\caption{Upper/lower bounds of the form factors $f^+$ for sample 1, 2, 3.}
\label{fig:fuplo_sample}
\end{figure}

In step 3, we compute the upper/lower bounds for the partial decay 
rate divided by $|V_{ub}|^2$ for each $q^2$ bins. To illustrate 
how it works, we pick up 3 generic set of samples 
( samples 1, 2 and 3 ). In Fig.~\ref{fig:fuplo_sample}
we show the upper/lower bounds of the form factors for each sample.  
It should be noted the open window for each sample from
$F^+_{up/lo}(q^2)$ is very small. 
\begin{figure}[here]
\resizebox{80mm}{!}{\includegraphics{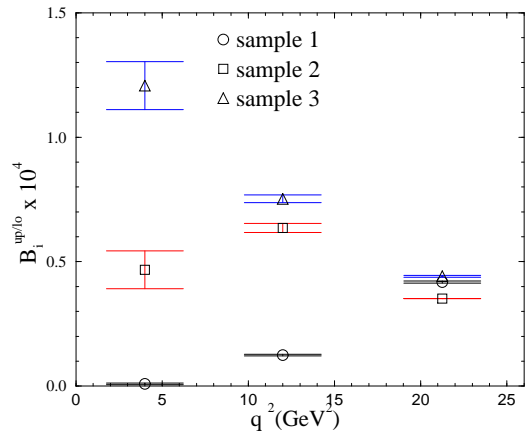}}
\caption{Upper/lower bounds of the partial branching fraction 
over the three energy bins for sample 1, 2, 3, 
where $B_i^{up/lo} \equiv \gamma_i^{up/lo}
|V_{ub}|^2/\Gamma_{total}$. While the overall normalization is 
unknown up to $|V_{ub}|$, we have set $|V_{ub}|=3.71\times 10^{-3}$ 
for illustration.}
\label{fig:gammasample}
\end{figure}
%
Integrating the squares of these upper/lower bounds for the 
form factors for each sample, we obtain the upper/lower bounds 
for $\gamma^{up/lo}_i$ and the partial branching fraction 
$B_i^{up/lo} \equiv \gamma_i^{up/lo}
|V_{ub}|^2/\Gamma_{total}$
for each $q^2$ bin as shown in Fig.~\ref{fig:gammasample}. 
For comparison we also show the the partial branching fraction 
by CLEO collaboration in Fig.~\ref{fig:Gamma_CLEO}.
%
\begin{figure}[here]
\includegraphics[width=9cm]{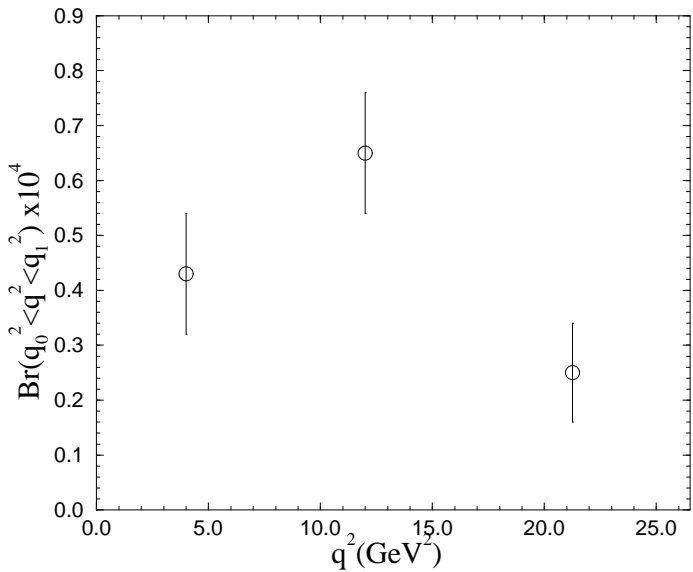}
\caption{The partial branching fraction $B_i$'s over the three energy 
bins from  CLEO, where  $B_i \equiv \Gamma_i/\Gamma_{total}$.}
\label{fig:Gamma_CLEO}
\end{figure}

In step 5, we impose condition B. 
As can be seen from 
Fig.~\ref{fig:gammasample} and Fig.~\ref{fig:Gamma_CLEO},
it is expected that the conditional probability with condition A+B 
would be highly suppressed for samples 1 and 3 but not so suppressed 
for sample 2.

\begin{figure}[here]
  \begin{center}
\resizebox{80mm}{!}{\includegraphics[angle=-90]{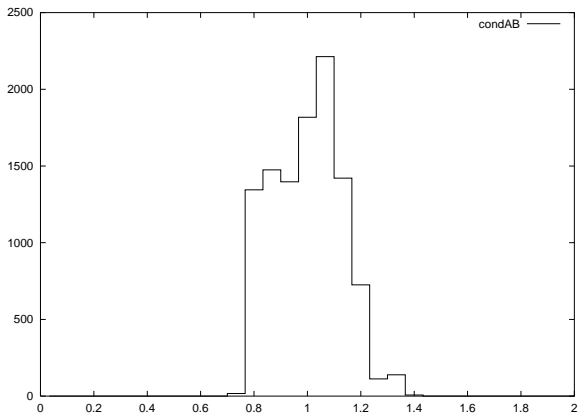}}
\caption{distribution of  $f^+(q^2=17.79GeV^2)$ 
from the samples which survived condition A and B .}
\label{fig:histvecfAB}
\end{center}
\end{figure}

\begin{figure}[here]
  \begin{center}
\resizebox{100mm}{!}{\includegraphics[angle=-90]{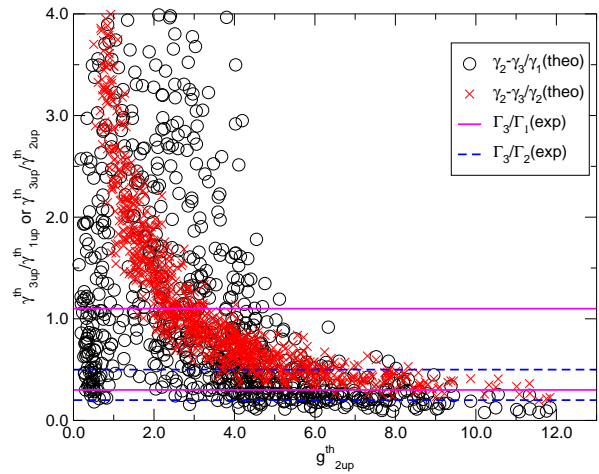}}
\caption{scatter plot of $\gamma^{th}_3/\gamma^{th}_2$ vs
$\gamma^{th}_2$ and $\gamma^{th}_1/\gamma^{th}_2$ vs $\gamma^{th}_2$ 
under condition A. As a reference we also show the favored region 
of $\Gamma^{exp}_3/\Gamma^{exp}_2$  and $\Gamma^{exp}_1/\Gamma^{exp}_2$  }
\label{fig:gam_corrA}
\end{center}
\end{figure}

The histogram for the samples of $\vec{f}$ after imposing 
condition B is shown in Fig.~\ref{fig:histvecfAB}. 
We find that the condition B reduces the variance 
of the form factor distribution even further.
%
In order to understand the mechanism of the reduction 
of the variance, 
%
we show show the scatter plot of $\gamma_3/\gamma_2$ vs $\gamma_2$
 or $\gamma_1/\gamma_2$ vs $\gamma_2$ in Fig.~\ref{fig:gam_corrA} .
We find that there are clear correlations.
%
From the CLEO data in Fig.~\ref{fig:Gamma_CLEO}, 
it can be shown that the experimental data favors 
the range $\Gamma^{exp}_3/\Gamma^{exp}_2 \sim 0.2 - 0.5$ and 
$\Gamma^{exp}_1/\Gamma^{exp}_2 \sim 0.4-0.9$. 
\begin{figure}[here]
  \begin{center}
\resizebox{100mm}{!}{\includegraphics[angle=-90]{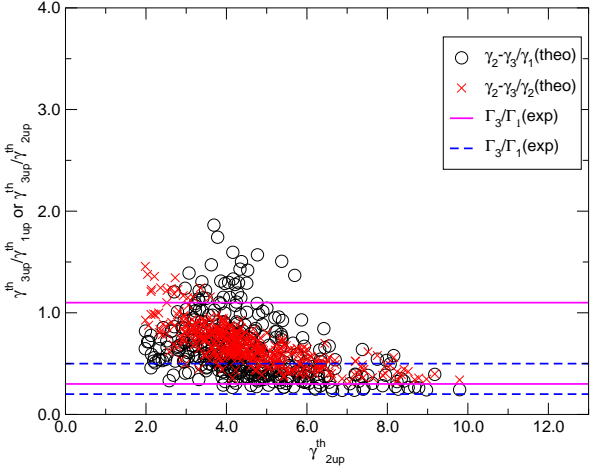}}
\caption{scatter plot of $\gamma^{th}_3/\gamma^{th}_2$ vs
$\gamma^{th}_2$ and $\gamma^{th}_1/\gamma^{th}_2$ vs $\gamma^{th}_2$ 
after imposing condition A+B.  As a reference we also show the 
favored region of $\Gamma^{exp}_3/\Gamma^{exp}_2$  and 
$\Gamma^{exp}_1/\Gamma^{exp}_2$.  }
\label{fig:gam_corrAB}.
\end{center}
\end{figure}
Since the open window of upper/lower bounds for each sample 
is quite small, imposing condition B is similar to restricting 
$\gamma_3/\gamma_2 = 0.2 - 0.5$ and $\gamma_1/\gamma_2 = 0.4 - 0.9$
in Fig.~\ref{fig:gam_corrA}. Since there is a strong correlation
this restricts $\gamma_2$. 
In fact after imposing condition B, the scatter plot 
corresponding to Fig.~\ref{fig:gam_corrA} looks 
as Fig.~\ref{fig:gam_corrAB}.

\subsection{results}

We give our final results for various physical quantities 
by looking their distributions which can be extracted from 
the conditional distribution $P_{AB}$. 

\begin{figure}[here]
\includegraphics[width=9cm]{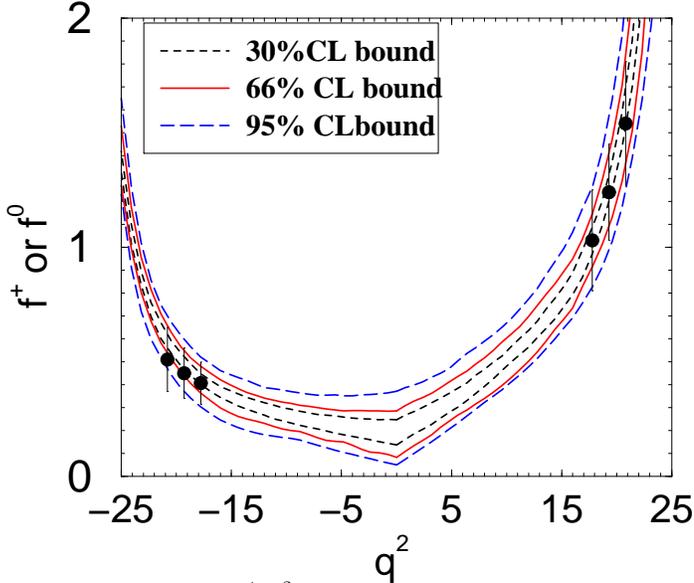}
\caption{CLB for $f^+(q^2)$. Here we used JLQCD's lattice input, 
and CLEO's experimental data.}
\label{fig:CLB_AB} 
\end{figure}

First, the dispersive bound is shown in Fig.~\ref{fig:CLB_AB}.
%
\begin{figure}[here]
\resizebox{80mm}{!}{\includegraphics[angle=-90]{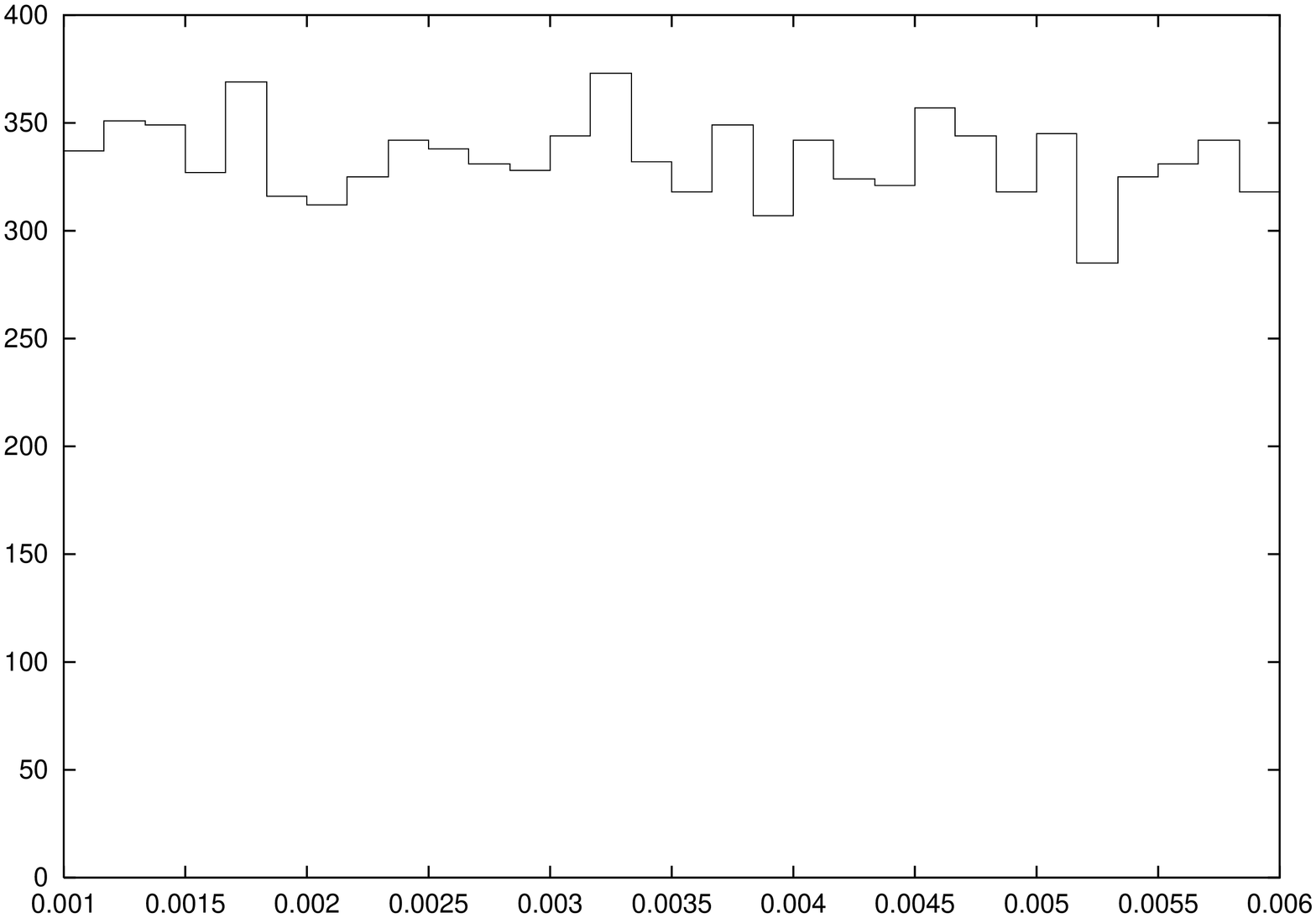}}
\resizebox{80mm}{!}{\includegraphics[angle=-90]{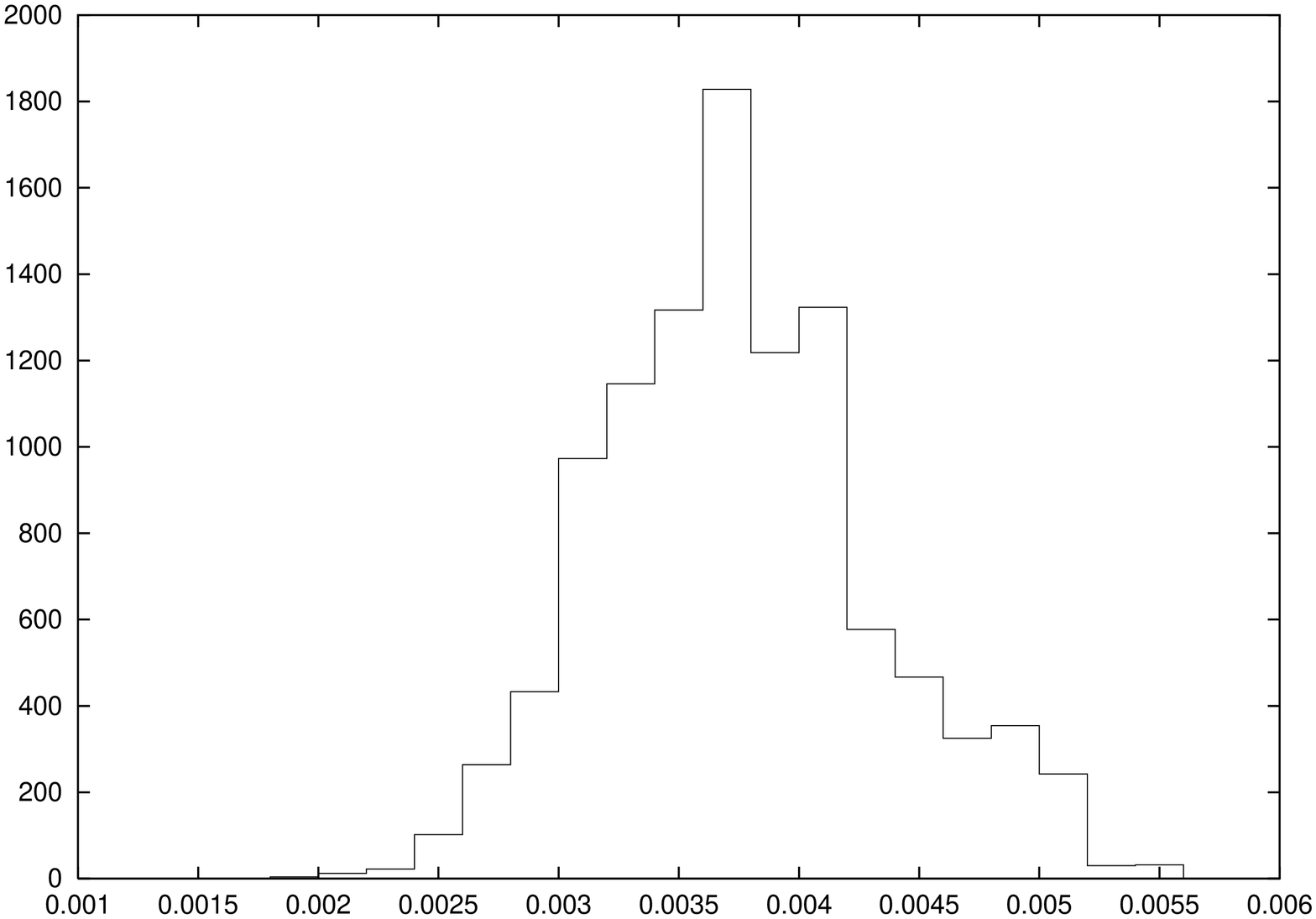}}
\caption{Histograms of $|V_{ub}|$. 
The original distribution was assumed to be flat in the 
range $|V_{ub}|=[1,6]\time  10^{-3}$ (left). 
After imposing condition B, the resulting distribution 
distributes with small variance.}
\label{fig:Vub_hist} 
\end{figure}
%
Fig.~\ref{fig:Vub_hist} shows the histogram of $|V_{ub}|$.
From this distribution we obtain 
\begin{eqnarray}
|V_{ub}| &=& [ 3.73 \pm 0.53 ] \times 10^{-3}.
\end{eqnarray}
This can be compared to the $|V_{ub}|$ determination 
using only the raw lattice data and the CLEO result at highest 
$q^2$ bin. For example, if we use the JLQCD results of the 
differential decay rate for $q^2 > 16$ GeV$^2$, 
we obtain $|V_{ub}| = [ 3.1\pm 0.9 ] \times 10^{-3}$.
This means that the $|V_{ub}|$ error of 30\% in the conventional 
method is reduced to 14 \% in our new method.

This error reduction is rather remarkable and requires some 
explanation on which part of the analysis contributed most. 
One important ingredient is the condition A and the other 
important ingredient is the condition B. Also, whether 
or not we have a soft pion input is another important point.
In order to see the effect of each ingredients, in addition 
to our full analysis we also carried out analyses with the 
following three cases.
\begin{enumerate}
\item Condition A only without soft pion input,
\item Condition A only and with soft pion input,
\item Condition A + B  but without soft pion input.
\end{enumerate}

\begin{figure}[t]
\resizebox{80mm}{!}{\includegraphics[angle=-90]{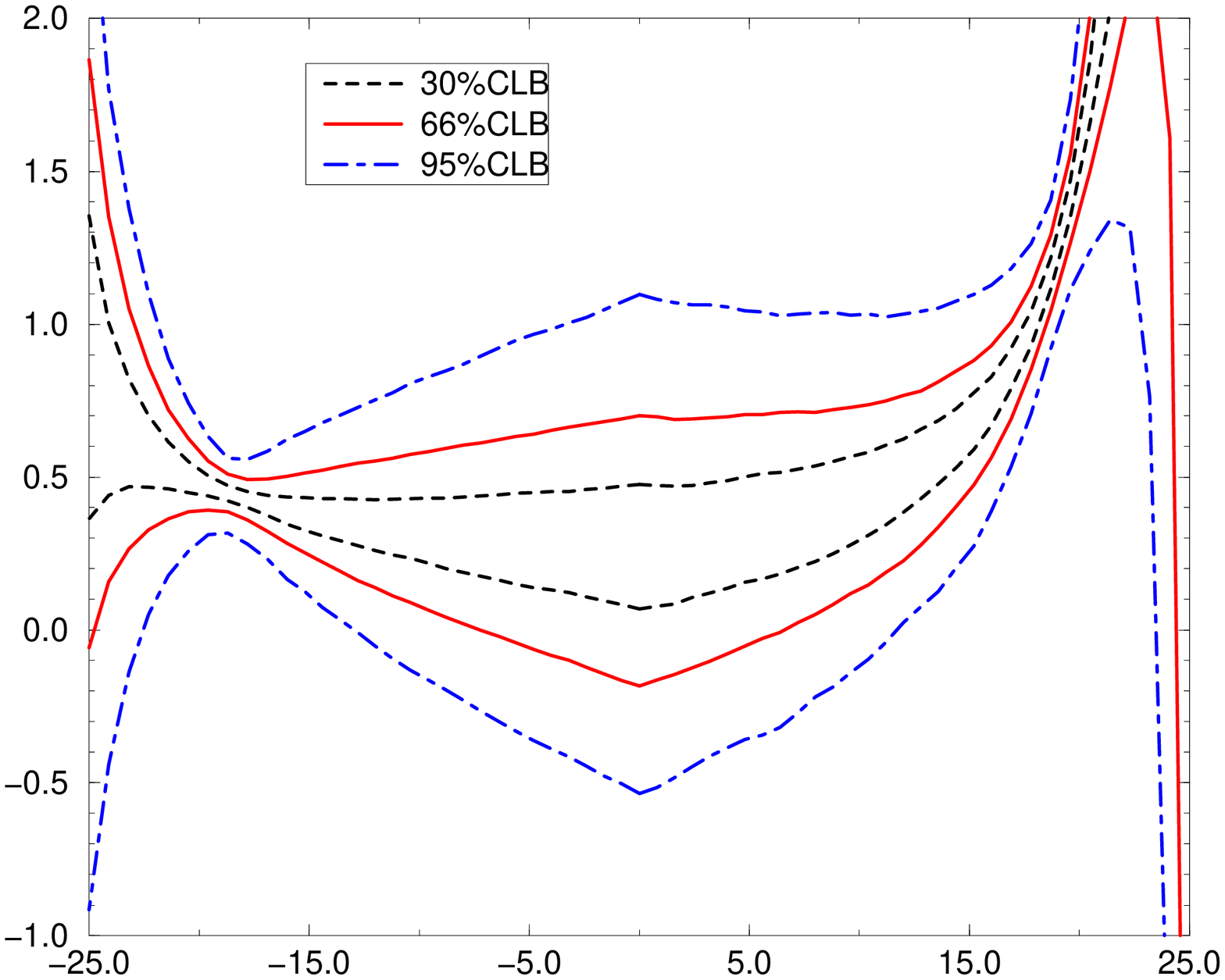}}
\caption{the case of Condition A only,  without soft pion input.}
\label{fig:A_noSPT}
\end{figure}

\begin{figure}[t]
\resizebox{80mm}{!}{\includegraphics[angle=-90]{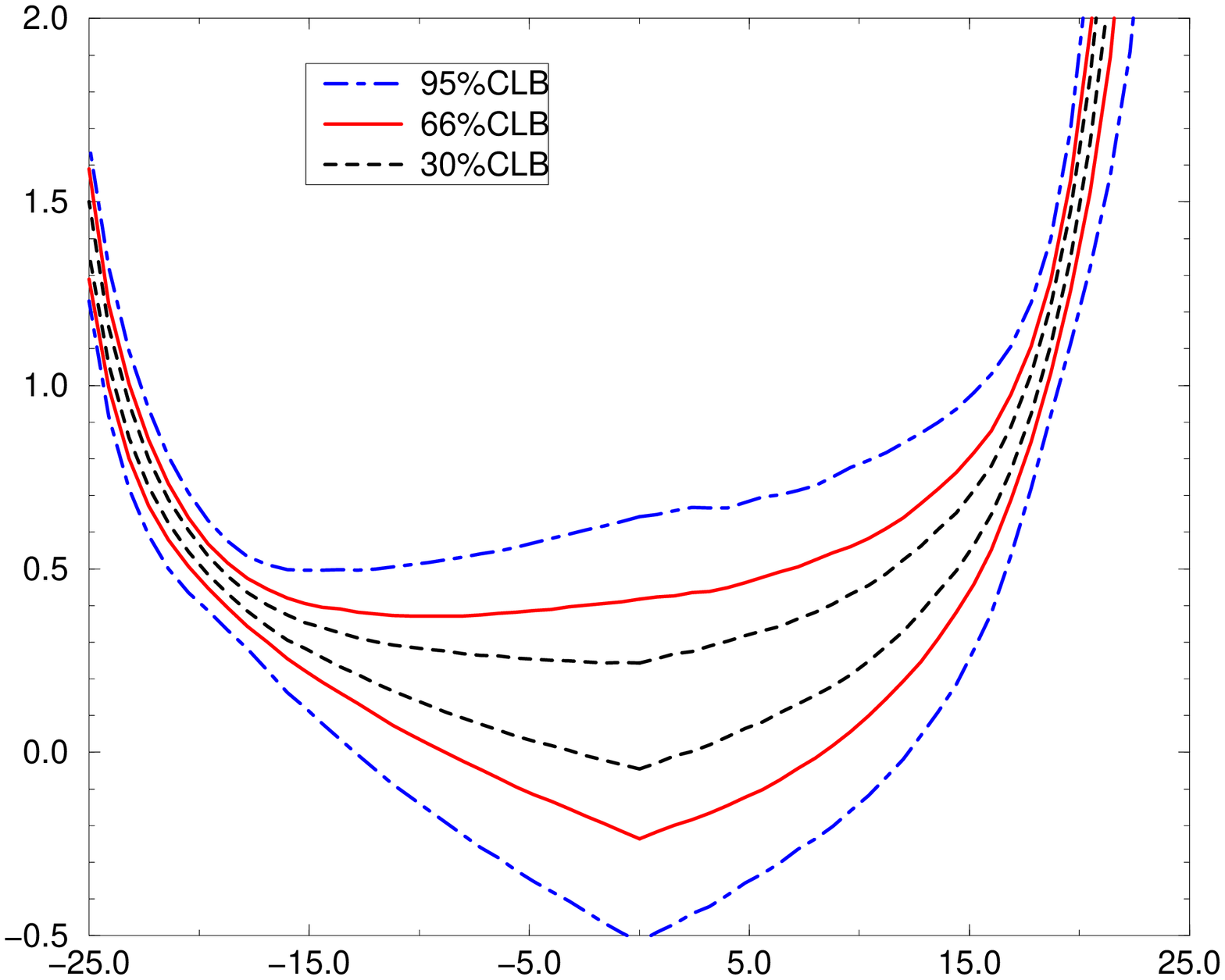}}
\caption{the case of Condition A only, with soft pion input.}
\label{fig:A_SPT}
\end{figure}
\begin{figure}[t]
\resizebox{80mm}{!}{\includegraphics[angle=-90]{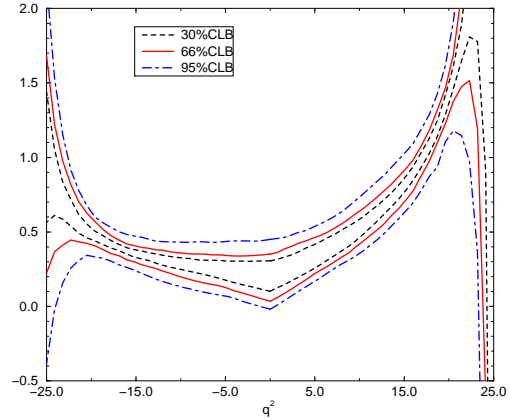}}
\caption{the case of Condition A+B,  without soft pion input.}
\label{fig:AB_noSPT}
\end{figure}

Fig.~\ref{fig:A_noSPT}, Fig.~\ref{fig:A_SPT}, and 
Fig.~\ref{fig:AB_noSPT} show the corresponding results.

The figures suggests that both condition A and condition B 
reduces the error. In fact, the corresponding $|V_{ub}|$ 
values are 
\begin{eqnarray}
|V_{ub}| &=& [ 2.92^{+0.65}_{-0.60} ]\times 10^{-3} \mbox{ for case 1},\\
|V_{ub}| &=& [ 2.99^{+0.63}_{-0.49} ]\times 10^{-3} \mbox{ for case 2},\\
|V_{ub}| &=& [ 3.61^{+0.55}_{-0.46} ]\times 10^{-3} \mbox{ for case 3},
\end{eqnarray}
so that the error for $|V_{ub}|$ is has 20\%, 20\% and 14\% 
for cases 1, 2 and 3 respectively. Thus we find both condition A 
and condition B reduces the error and having both of them together 
gives the significant reduction. It is also found that the input from 
the soft pion theorem does not change the result of $|V_{ub}|$ so much.

We also obtain  product we also obtain 
bounds  at 66\% confidence level (66\% CL) for 
 $f^+(0)$ as a byproduct
\begin{eqnarray}
0.126< f(0)< 0.293.
\end{eqnarray}
which will be very useful for predicting the two body decay rate 
in QCD factorization .

\section{Systematic Errors}
\label{sec:Errors}
In this section we discuss possible systematic errors 
in our analysis.

The main systematic error arise from the lattice data 
and from the experimental data. These errors are of course 
already included in the errors of our input data. However, 
it would be nice if we could estimate these errors in a 
different fashion. One way to do so is to use the input data from 
other groups which has different systematic errors. 
In the following we present our result using different lattice inputs. 
Although we should also perform analyses using other 
experimental data such as those from Belle or BaBar, 
this will be left as a future problem since there is no
publicly available data. Therefore we will not discuss 
the systematic errors from experimental data in this paper.

There are four quenched lattice calculations with different 
systematic errors. JLQCD collaboration used nonrelativistic 
effective theory for the heavy quark on a coarse lattice, 
in which the dominant systematic errors are the momentum dependent 
discretization error and chiral extrapolation error. Fermilab 
collaboration also used a different type of nonrelativistic 
effective theory on relatively coarse lattices, using different 
chiral extrapolation method.
APE collaboration and UKQCD collaboration used relativistic action 
for the heavy quark on finer lattices, in which the dominant 
error is the presumably momentum independent discretization error 
from the large heavy quark mass in lattice units.
In Table~\ref{tab:lattice}, we give the input parameters from 
APE, UKQCD and Fermilab collaborations.
 \begin{table}[here]
 \begin{center}
  \begin{tabular}{|c|c|c|c|}           \hline
 collaboration & $q^2(GeV^2)$ & $f^+(q^2)$      & $f^0(q^2)$    \\ \hline
 Fermilab
      & $17.23$      & $1.13^{+0.29}_{-0.19}$  & $0.64^{+0.13}_{-0.10}$ \\
      & $20.35$      & $1.72^{+0.31}_{-0.27}$  & $0.83^{+0.16}_{-0.13}$ \\
      & $23.41$      & $2.10^{+0.43}_{-0.40}$  & $1.00^{+0.20}_{-0.15}$ \\
\hline
 APE  & $13.6$      & $0.70^{+0.13}_{-0.09}$  & $0.46^{+0.09}_{-0.11}$ \\
      & $17.9$      & $1.05^{+0.15}_{-0.13}$  & $0.59^{+0.07}_{-0.11}$ \\
      & $22.1$      & $1.96^{+0.24}_{-0.30}$  & $0.80^{+0.06}_{-0.13}$ \\
\hline
 UKQCD& $16.7$      & $0.90 \pm 0.22$         & $0.57^{+0.06}_{-0.11}$ \\
      & $18.1$      & $1.10^{+0.28}_{-0.22}$  & $0.61^{+0.06}_{-0.11}$ \\
      & $22.3$      & $2.30^{+0.67}_{-0.36}$  & $0.79^{+0.05}_{-0.11}$ \\
\hline
  \end{tabular}
  \caption{lattice results of $f^+(q^2)$ and $f^0(q^2)$ by 
   Fermilab, APE and UKQCD.}
\label{tab:lattice}
 \end{center}
\end{table}
The setups for the numerical analysis are almost the same 
except for the number of generated samples for $\{\vec{f},g,f_B\}$
in step 1. In the analyses using Fermilab, APE and UKQCD data,  
we have generated ${\cal O}(10^5)$ samples.
The results using Fermilab data are
\begin{eqnarray}
|V_{ub}| = [ 3.59^{+0.42}_{-0.40} ] \times 10^{-3}, \\
0.128 < f(0)< 0.339 \mbox{  (at  66\% CL)}, 
\end{eqnarray}
those using APE data are
\begin{eqnarray}
|V_{ub}| = [ 3.35^{+0.45}_{-0.49} ] \times 10^{-3}, \\
0.132 < f(0)< 0.302 \mbox{   (at 66\% CL)}, 
\end{eqnarray}
those using UKQCD data are
\begin{eqnarray}
|V_{ub}| = [ 3.37^{+0.46}_{-0.49} ] \times 10^{-3}, \\
0.154 < f(0)< 0.353 \mbox{  (at  66\% CL)} .
\end{eqnarray}

Another possible source of systematic errors are 
the theoretical input values of $\chi_L(0)$, $\chi_T(0)$ for 
the dispersive bound which is obtained by the operator product 
expansion (OPE) using perturbative QCD and some estimate of 
vacuum condensation values. (See appendix for more details.) 
As is explained in the appendix these input values are 
\begin{eqnarray}
\chi_T &=& [5.60 \pm 0.17]\times 10^{-4} ,\\
\chi_L &=& [1.50 \pm 0.03]\times 10^{-2} ,
\end{eqnarray}
where the errors are from unknown 2-loop perturbative corrections
and uncertainties in the vacuum condensation values.
By changing the input values of $\chi_T$ and $\chi_L$ 
by 1 $\sigma$, the central values of 
$|V_{ub}|$ changes by 2\% when JLQCD data are used, which is negligible
compared to other errors. We also found that the same is true
when Fermilab, APE, or UKQCD data are used.

Yet another systematic error is the quenching error in the lattice
results. Since there is no way to control this error other than 
performing unquenched lattice calculations, which is beyond the 
scope of this paper. We leave it an open question and wait for 
the unquenched lattice results, which is expected to appear near
future.

\section{Conclusion}
\label{sec:Conclusion}
In this paper, we have proposed a method to determine 
$|V_{ub}|$ by combining lattice results, dispersive bounds, 
and experimental data. 
Based on Lellouch's idea we considered the statistical distributions 
of the form factors and their bounds. Our proposal is to restrict the 
distribution by imposing a physical condition using the experimental 
data. This gives a significant reduction in the errors of $|V_{ub}|$ .
As as result we obtained 
\begin{eqnarray}
|V_{ub}| &=& [ 3.73 \pm 0.53 ] \times 10^{-3}.
\end{eqnarray}
As a by product we also obtained a 66\% confidence level bound 
for the form factor at small $q^2$,
\begin{eqnarray}
0.293&< f(0)< 0.126
\end{eqnarray}
which will be very useful for predicting the two body decay rate 
in QCD factorization. 

The recent Belle results of $|V_{ub}|$ from $B\rightarrow X_u l\nu$~
\cite{Kakuno:2003fk}is
\begin{equation}
|V_{ub}| = [ 4.66\pm 0.28\pm 0.35\pm 0.17\pm 0.08\pm 0.58]\times 10^{-3},
\end{equation}
where the errors is the statistical, systematic, $b\rightarrow c$,
$b\rightarrow u$, and theoretical.
The lattice QCD based $|V_{ub}|$ values by 
CLEO collaboration from $B\rightarrow\pi l\nu$ is 
\begin{eqnarray}
|V_{ub}| &=& [ 2.88  \pm 0.55 \pm 0.30 {}^{+0.45}_{-0.35}\pm 0.18 ] 
\times 10^{-3},
\end{eqnarray}
where the errors are statistical, experimental systematic,
theoretical and $\rho l\nu$ form factor shape.
It should be noted that this discrepancies are reduced 
by our method although the input lattice data are the same.

For future studies, we should elaborate our method in such a way 
that we treat the systematic and statistical errors separately 
which would give a more careful analysis of errors. This is 
in principle straightforward but requires more numerical calculation
since the generation of random samples should be much increased. 

Another improvement is to update the input data of form factors 
from lattice, especially the unquenched one, and the input data 
of partial decay rates from B factories such as Belle and BaBar.
Both of these will be available near future. 
 
\section*{Appendix: Review of the Dispersive Bound}
\label{sec:DB}
In this appendix, we briefly summarize the dispersive bounds on the form 
factors, which is fully discussed in Ref.~\cite{Lellouch:1995yv}. 
The matrix element of the heavy-to-light semileptonic decay
$B\rightarrow \pi l\nu$ is parameterized as Eq.~(\ref{eqff}).
The form factors $f^+$ and $f^0$ satisfies the following kinematical 
constraint
\begin{equation}
f^+(0)=f^0(0). 
\label{eqkin}
\end{equation}
In order to derive bound on $f^+(q^2)$ and $f^0(q^2)$ we consider 
the following two-point function
 \begin{eqnarray}
 \Pi^{\Gm\Gn}(q)
  &\equiv& i\int d^4x \e^{iqx}\bra{0}TV^\Gm(x)V^\Gn(0)\ket{0}
\nn\\
  &=&-(g^{\Gm\Gn}q^2-q^\Gm q^\Gn)\Pi_T(q^2)+q^\Gm q^\Gn\Pi_L(q^2) ,
\label{eqPi} 
 \end{eqnarray}
where $\Pi_{T(L)}$ corresponds to the contribution from the 
intermediate states with $J^P=1^-(0^+)$. In the deep Euclidean region 
which is far from the physical cut, this two point function can be 
evaluated by perturbative QCD reliably. 

Inserting all possible intermediate states $|\Gamma\rangle$ we obtain 
the following result for the imaginary part of $\Pi_{\Gm\Gn}$:
\begin{eqnarray}
& &  -(g^{\Gm\Gn} q^2-q^\Gm q^\Gn){\rm Im}\Pi_T+q^\Gm q^\Gn {\rm Im}\Pi_L
\nn\\
&=& 
\frac{1}{2}\sum_\Gamma(2\pi)^4\Gd^4(q-p_\Gamma)
\bra{0}V^\Gm\ket{\Gamma}\bra{\Gamma}V^{\Gn^\dagger}\ket{0},
\label{eqImPi}
 \end{eqnarray}
We now restrict ourselves to include only contributions of the $B\pi$
and the $B^*$ states. Since each hadron state contributes positively to 
the spectral function, we derive the following two bounds for the 
form factors
\begin{eqnarray}
  {\rm Im}\Pi_L(t)&\ge&
\frac{3 t_+ t_-}{32\pi}\sqrt{(t-t_+)(t-t_-)
}\frac{|f^0(t)|^2}{t^3}\Gt(t-t_+),
\label{eqPiL}
\nn\\
{\rm Im}\Pi_T(t)
&\ge &\pi\left(\frac{m_{B^*}}{f_{B^*}}\right)^2\Gd(t-{m_{B^*}}^2)
\nn\\
& &
+\frac{1}{32\pi}\frac{[(t-t_+)(t-t_-)]^{3/2}}{t^3}
|f^+(t)|^2\Gt(t-t_+),
\nn
\label{eqPiT}
 \end{eqnarray}
where $t\equiv q^2$ and $t_{\pm}\equiv (m_B\pm m_\pi)^2$ . 
In the above equations we have limited ourselves to contribution of the 
$B^*$ state with 
\begin{equation}
  \bra{0}V^\Gm\ket{B^*(r,p)}=\frac{m_{B^*}^2}{f_{B^*}}\Ge^\Gm_r
\end{equation} 
  In QCD, the polarization functions $\Pi_L$ and $\Pi_T$ obey once and 
twice subtracted dispersion relation($q^2=-Q^2$), i.e.:  
 \begin{eqnarray}
\chi_L(Q^2)
&\equiv&
-\frac{\partial}{\partial Q^2}(-Q^2\Pi_L(Q^2))
\nn\\
&=& \int^\infty_0\frac{dt}{\pi}\frac{t{\rm Im}\Pi_L(t)}{(t+Q^2)^2} 
\label{eqchiL}\\
\chi_T(Q^2)
&\equiv&
\frac{1}{2}\left(-\frac{\partial}{\partial
Q^2}\right)^2(-Q^2\Pi_T(Q^2))
\nn\\
& & 
=\int^\infty_0\frac{dt}{\pi}\frac{t{\rm Im}\Pi_T(t)}{(t+Q^2)^3} 
\label{eqchiT}
 \end{eqnarray}
Thus it is now possible to get bounds on the form factors to inserts
 Eqs.(\ref{eqchiL},\ref{eqchiT}) to Eqs.(\ref{eqPiL},\ref{eqPiT})
 \begin{eqnarray}
  J_{L}(Q^2)
&\ge& 
\frac{1}{\pi}\int_{t_+}^\infty dt k_{L}(t,Q^2)|f_{0}(t)|^2
\label{eqJL}\\
  J_{T}(Q^2)
&\ge& 
\frac{1}{\pi}\int_{t_+}^\infty dt k_{T}(t,Q^2)|f_{+}(t)|^2
\label{eqJT}
 \end{eqnarray}
where $J_{L(T)},k_{L(T)}$ are
 \begin{eqnarray}
  J_L&=&\chi_L(Q^2)
\label{eqJL2},\\
  J_T&=&\chi_T(Q^2)-\frac{1}{(m_{B^*}+Q^2)^3}
  \left(\frac{m_{B^*}^2}{f_{B^*}}\right)^2
\label{eqJT2},\\
  k_L&=&\frac{3}{32\pi}\frac{[(t-t_+)(t-t_-)]^{1/2}t_+t_-}{(t+Q^2)^2t^2},
\\
  k_T&=&\frac{3}{96\pi}\frac{[(t-t_+)(t-t_-)]^{3/2}}{(t+Q^2)t^2}
 .
 \end{eqnarray}
To obtain the bounds of the form factor for values of t in the range 
$[0,t_-]$, we map the complex t-plane into the unit disc in complex
z-plane with the conformal transformation
 \begin{equation}
  \frac{1+z}{1-z}=\sqrt{\frac{(t_+-t)(t_--t)}{t_+-t_-}}.\nn
 \end{equation}
This maps the contour along the physical cut 
in Eqs.(\ref{eqJL},\ref{eqJT}) onto a unit circle 
in z-plane so that 
  \begin{eqnarray}
   J_L(Q^2)&\ge& \oint_c\frac{dz}{2\pi iz}|\phi_L(z,Q^2)f_0(z)|^2,
\label{eqJL3}\\
   J_T(Q^2)&\ge& \oint_c\frac{dz}{2\pi iz}|\phi_T(z,Q^2)f_+(z)|^2,
\label{eqJT3}
  \end{eqnarray}
where we have used the fact that $k(t,Q^2)$ is positive definite
quantity. The functions $\phi_{L(T)}$ are defined as 
\begin{eqnarray}
\phi_L
&=&
\sqrt{\frac{3t_+t_-}{4\pi}}\frac{1}{t_+-t_-}\frac{1+z}{(1-z)^{5/2}}
\left(\sqrt{\frac{t_+}{t_+-t_-}}+\frac{1+z}{1-z}\right)^{-2}
\nn\\
& &
\left(\sqrt{\frac{t_+-Q^2}{t_+-t_-}}+\frac{1+z}{1-z}\right)^{-2},
\label{eqphiL} \nn \\ 
\phi_T
&=&
\sqrt{\frac{1}{\pi(t_+-t_-)}}\frac{(1+z)^2}{(1-z)^{9/2}}
\left(\sqrt{\frac{t_+}{t_+-t_-}}+\frac{1+z}{1-z}\right)^{-2}
\nn\\
&& \left(\sqrt{\frac{t_+-Q^2}{t_+-t_-}}+\frac{1+z}{1-z}\right)^{-3},
\label{eqphiT} \nn 
\end{eqnarray}
and satisfy $\phi(z,Q^2)\ge 0$ for $t\in[0,t^-]$.  
For simplicity, we define an inner product on the unit circle 
\begin{equation}
 \bk{g}{h}=\oint_{|z|=1}\frac{dz}{2\pi iz}g^*(z)h(z),
\end{equation}
so that the inequality Eq.~(\ref{eqJL3}), Eq.~(\ref{eqJT3}) can be written 
\begin{equation}
 J \ge \bk{\phi f}{\phi f}.
\end{equation}

 And define the function $g_t(z)$ and the matrix $M(f(t))$ as
\begin{eqnarray}
g_t(z) &\equiv & \frac{1}{1-z^*(t)z},\nn\\
M(f(t))&=&\begin{pmatrix}
	\bk{\phi f}{\phi f} & \bk{\phi f}{g_t}\\
	\bk{g_t}{\phi f }  & \bk{g_t}{g_t},
	\end{pmatrix}\nn
\end{eqnarray}
then if $f(z(t))$ has no pole in the range $[0,t_-]$, 
$f(z(t))$ can be related the inner-product of $g_t$ and 
$\phi f$ as 
\begin{equation}
 \bk{g_t}{\phi f} = \phi(z(t),Q^2)f(z(t)) 
\label{eqgt_phif}
\end{equation}
Now, because positivity of inner products  
\begin{equation}
 \det M(f(t),f(t_1)) \ge 0
\label{eqdetM}
\end{equation}
is  satisfied, by eliminating $\bk{\phi f}{\phi f}$ with 
(\ref{eqJL3},\ref{eqJT3}) we get the form factors bound as  
\begin{equation}
 |f(t)|^2\le J(Q^2)\frac{1}{1-|z(t)|^2}\frac{1}{\phi(z(t),Q^2)}.
\label{eqff_bound}
\end{equation}
If we want to get  more conditionality bound, using some values 
of the form factors at $t_1,t_2,\cdots,t_L$ and defining the matrix 
$M(f(t),\vec{f})$ as
\begin{equation}
 M(f(t),\vec{f})=
  \begin{pmatrix} 
   \bk{\phi f}{\phi f} & \bk{\phi f}{g_t} & \bk{\phi f}{g_{t_3}} 
   & \cdots & \bk{\phi f}{g_{t_N}} \\
   \bk{g_t}{\phi f}    & \bk{g_t}{g_t}    & \bk{g_t}{g_{t_3}}    
   & \cdots & \bk{g_t}{g_{t_N}}    \\
   \bk{g_{t_3}}{\phi f}& \bk{g_{t_3}}{g_t}& \bk{g_{t_3}}{g_{t_3}}
   & \cdots & \bk{g_{t_3}}{g_{t_N}}\\
   \vdots              & \vdots           & \vdots               
   & \vdots & \vdots               \\
   \bk{g_{t_N}}{\phi f}& \bk{g_{t_N}}{g_t}& \bk{g_{t_N}}{g_{t_3}} 
   & \cdots & \bk{g_{t_N}}{g_{t_N}}
  \end{pmatrix},
\label{eqdefM}
\end{equation}
where $\vec{f}$ stand for $\{f(z(t_1)),f(z(t_2)),\cdots,f(z(t_L))\}$.
This matrix $M(f(t),\vec{f})$ satisfies
\begin{equation}
det M(f(t),\vec{f}) \ge 0\nn, 
\label{eqdetM2}
\end{equation}
By eliminating $\langle \phi f | \phi f \rangle$ 
using (\ref{eqJL3},\ref{eqJT3}), the inequality~(\ref{eqdetM2}) leads 
to bounds on the form factors $f(t)$, 
\begin{equation}
 F_{lo}(t;\vec{f})\le f(t) \le F_{up}(t;\vec{f}). \nn
\end{equation}
This upper and lower functions $F_{up(lo)}$ stand for 
\begin{equation}
 F_{up/lo}=\frac{-\beta(t)+/- \sqrt{c(Q^2)\cdot\Delta(t)}}
{\alpha\cdot \phi(t,Q^2)}, 
\label{eqff_bound2}
\end{equation}
where $\alpha$, $\beta(t)$, and $\sqrt{c(Q^2)\cdot\Delta(t)}$ 
are known functions of $Q^2$, $t_i$, and $f(z(t_i))$ ($i=1,\cdots,L$).
The particular form of the Eqs.~(\ref{eqff_bound2}) can be found 
in Ref.~\cite{Lellouch:1995yv}.

 Now, if $f_+(t)$ has a single pole at $t=t_p$ away from the cut 
(in fact $t_p=m_{B^*}\in [t_-,t_+]$), Eqs.~(\ref{eqgt_phif}) becomes       

\begin{equation}
\bk{g_t}{\phi f}
=\phi(z(t))f(z(t))+\frac{Res(\phi f;z(t_p))}{z(t_p)-z(t)}\nn 
\end{equation}
so $\bk{g_t}{\phi f}$ have the single pole at $z=z(t_p)$.
In this case we define the function $\phi_p(z)$, which satisfies
$|\phi_p|=|\phi|$ at $|z|=1$, as, 
\begin{equation}
 \phi_p(z,Q^2)\equiv \phi(z,Q^2)\frac{z-z(t_p)}{1-z^*(t_p)z}. \nn
\end{equation}
This function $\phi_p f_p$ dose not have a pole at $z=z(t_p)$, so  
\begin{eqnarray}
 \bk{\phi_p f_p}{g_t}&=&\phi_p(t) f_p(t)\nn\\
 \bk{\phi_p f_p}{\phi_p f_p}&=&\bk{\phi f}{\phi f}\le J(Q^2)\nn
\end{eqnarray}
are satisfied. Thus, the bound of Eqs.~(\ref{eqff_bound2})holds
\begin{eqnarray}
 F_{lo}^p(t;\vec{f})\le f(t) \le F_{up}^p(t;\vec{f}).\nn
\end{eqnarray}
where $F_{lo,up}^p$ are the functions $F_{lo,up}$ obtained by
replacing $\phi$ with $\phi_p$ . In this paper 
we denote $\phi_p,F_{lo,up}^p$  as  $\phi,F_{lo,up}$ for $f^+$.
In this way, we get the the independent bounds on the form factors 
using the given values of the discrete set of points 
,for example top figure of Fig.~\ref{fig:disp_no_error} 
with no kinematical constraint. But form factors not independent but 
satisfy the kinematical constraint Eqs.\ref{eqkin}. So we make a new
dispersive bound, use 
\begin{equation}
f^+(0)_{up}=F_{up}^0(0,\vec{f}) 
\end{equation}
as the additional input of the new bound for $f^+(t)$,
and consider this upper bound as new appear bound for $f^+(t)$.
And equally make a new lower bound for $f^+(t)$, using the additional input 
\begin{equation}
f^+(0)_{lo}=F_{lo}^+(0,\vec{f}) .
\end{equation}
So we can make kinematical constraint dispersive bound like bottoms of
fig\ref{fig:disp_no_error}. There may exist 4 pattern no kinematical constraint
bound, from there kinematical constraint bound can be made
respectively same method.        
 
 In this calculation, they use OPE and calculate the
coefficient by perturbative QCD. 
In this paper, we take $\chi_T(0)=(5.60\pm 0.17)\times 10^{-4}, 
\chi_L(0)=(1.50\pm 0.03)\times 10^{-2}$
(See Ref.~\cite{Lellouch:1995yv}).
The values of $\chi$ in $\overline{MS}$ scheme are given 
in Table.~\ref{tab:chi}. 
These coefficients are obtained 
from the operator product expansion of hadron current correlators
in perturbative approximation.  
In order to control the higher order contributions which grow
for larger $-Q^2$, we take $Q^2=0$. 
The table shows that the $O(\alpha_s)/1Loop$ are 17\%,7\% for $\chi_T,
\chi_L$ so that the perturbative expansion is under control. 
The unknown 2-loop errors can be estimated by either naive order 
counting or by comparing results with different renormalization 
schemes, which amount to less than or equal to 1\%. 
We therefore estimate the uncertainties in $\chi$ from the 
condensates, which are of order 2\% and 3\% for $\chi_L$ and $\chi_T$.  
\begin{table}[here]
 \begin{center}
  \begin{tabular}{|c|c|c|}\hline        
  \multicolumn{1}{|c|}{}    
  & $\chi_L(Q^2=0)$ & $\chi_T(Q^2=0)$\\ \hline
  \multicolumn{1}{|c|}{$1Loop$}     
  & $1.3\times 10^{-2}$ &$5.1\times 10^{-4}$\\  \hline
  \multicolumn{1}{|c|}{$O(\alpha_s)/1Loop$} 
  & $17\%$ &$7\%$\\ \hline
  \multicolumn{1}{|c|}{total pert} 
 & $1.5\times10^{-2}$&$5.5\times 10^{-4}$ \\ \hline
  \multicolumn{1}{|c|}{$m_u<\bar{u}u>/1Loop$} 
 & $-2\%$ &$3\%$ \\ \hline
  \multicolumn{1}{|c|}{$<\alpha_sG^2>/1Loop$} 
 & $0.04\%$ &$-0.05\%$ \\ \hline
  \multicolumn{1}{|c|}{$total$} 
 & $1.5\times10^{-2}$&$5.6\times 10^{-4}$ \\ \hline
  \end{tabular}
  \caption{perturbative and non-perturbative contributions to
  subtracted polarization function $\chi_L$ and $\chi_T$ by
  $\bar{MS}$-scheme.
 ``1 Loop'' stands for the 1-loop result and ``totalpart'' 
  stands for  the results through $O(\alpha_s)$. 
$m_u<\bar{u}u>/1loop$,$ <\alpha_sG^2>$ are the higher twist 
contributions from the condensates. }
\label{tab:chi} 
 \end{center}
\end{table}

\section*{Acknowledgments}\label{sec:acknowlegments}

We acknowledge Shoji~Hashimoto, Laurent~Lellouch, Tohru~Iijima, 
and Takashi~Hokuue for stimulating discussions. We also would like to 
thank Hideo Matsufuru for his useful comments. Special thanks to 
Masanori Okawa for his reading of the manuscript and his comments.
We would also like to thank the Yukawa Institute for Theoretical 
Physics at Kyoto 
University, where we benefited from the discussions during the 
YITP-W-02-20 workshop on "QCD for B decays''.
T.~O. is supported by Grant-in-Aid for Scientific research 
from the Ministry of Education, 
Culture, Sports, Science and Technology of Japan (Nos. 13135213,
16028210, 16540243).

\end{document}